\documentclass[reprint,prd,aps,a4paper,superscriptaddress]{revtex4-1}

\usepackage{amssymb}
\usepackage{amsmath}
\usepackage{bbold}
\usepackage{color}
\usepackage{graphicx}
\usepackage{hyperref}
\usepackage{afterpage}
\usepackage{tabularx}
\usepackage{booktabs}
\usepackage{xspace}

\newcommand{\lcdm}{\ensuremath{\Lambda{\rm CDM}}\xspace}
\newcommand{\ere}{\langle D \rangle}

\begin{document}

\author{Sebastian Seehars}
\email{sebastian.seehars@phys.ethz.ch}
\affiliation{ETH Zurich, Department of Physics, Wolfgang-Pauli-Str. 27, 8093 Zurich, Switzerland}
\author{Sebastian Grandis}
\email{grandis@usm.lmu.de}
\affiliation{ETH Zurich, Department of Physics, Wolfgang-Pauli-Str. 27, 8093 Zurich, Switzerland}
\affiliation{Faculty of Physics, Ludwig-Maximilian University, Scheinerstr. 1, 81679 Munich, Germany}
\affiliation{Excellence Cluster Universe, Boltzmannstr. 2, 85748 Garching, Germany}
\author{Adam Amara}
\email{adam.amara@phys.ethz.ch}
\affiliation{ETH Zurich, Department of Physics, Wolfgang-Pauli-Str. 27, 8093 Zurich, Switzerland}
\author{Alexandre Refregier}
\email{alexandre.refregier@phys.ethz.ch}
\affiliation{ETH Zurich, Department of Physics, Wolfgang-Pauli-Str. 27, 8093 Zurich, Switzerland}

\title{Quantifying Concordance in Cosmology}

\begin{abstract}
Quantifying the concordance between different cosmological experiments is important for testing the validity of theoretical models and systematics in the observations. In earlier work, we thus proposed the Surprise, a concordance measure derived from the relative entropy between posterior distributions. We revisit the properties of the Surprise and describe how it provides a general, versatile, and robust measure for the agreement between datasets. We also compare it to other measures of concordance that have been proposed for cosmology. As an application, we extend our earlier analysis and use the Surprise to quantify the agreement between WMAP~9, Planck~13 and Planck~15 constraints on the $\Lambda$CDM model. Using a principle component analysis in parameter space, we find that the large Surprise between WMAP~9 and Planck~13 ($S = 17.6$ bits, implying a deviation from consistency at $99.8\%$ confidence) is due to a shift along a direction that is dominated by the amplitude of the power spectrum. The Planck~15 constraints deviate from the Planck~13 results ($S = 56.3$ bits), primarily due to a shift in the same direction. The Surprise between WMAP and Planck consequently disappears when moving to Planck~15 ($S = -5.1$ bits). This means that, unlike Planck~13, Planck~15 is not in tension with WMAP~9. These results illustrate the advantages of the relative entropy and the Surprise for quantifying the disagreement between cosmological experiments and more generally as an information metric for cosmology.
\end{abstract}

\maketitle

\section{Introduction}
\label{sec:introduction}

The current standard model of cosmology, a flat universe dominated by dark matter and dark energy (\lcdm), is in good agreement with a wealth of cosmological probes. As we collect and analyze ever more precise cosmological data, an important question is whether or not the different available probes are consistent with each other within a cosmological model. 

For this purpose, different measures of agreement between cosmological datasets have been introduced~\cite{2002MNRAS.335..377H,Marshall:2006hw,Amendola:2013ce,2014MNRAS.439.1855H,2014PhRvD..90f3501M,2015MNRAS.449.2405K,2015arXiv151000688R,Verde:2013hp,MacCrann:2015ke}. In~\cite{Seehars:2014ir}, we proposed a new measure that we called Surprise. It is derived from the relative entropy, or Kullback-Leibler divergence~\cite{Kullback:1951va}, between two posteriors and is a global measure of consistency over the entire cosmological parameter space. In~\cite{Seehars:2014ir}, we also applied the Surprise to a historical sequence of CMB experiments. One of the key results was a large Surprise when comparing the constraints from observations of the Wilkinson Microwave Anisotropy Probe (WMAP)~\cite{Bennett:2012wy, Hinshaw:2013dd} to the 2013 release of Planck data~\cite{PlanckCollaboration:2014km, PlanckCollaboration:2014kt}. In further work~\cite{Grandis:2015}, we applied the Surprise measure to a wide range of other cosmological probes and quantified their relative information gains and tensions between them.

In this paper, we revisit the properties of the Surprise and compare it to other proposed measures of agreement between cosmological datasets. 
We derive expressions for these different measures in the limit of Gaussian likelihoods and a linear model and illustrate the results using a one-dimensional toy model. As an application of the Surprise, we reexamine the consistency of Planck and WMAP in light of the 2015 data release from Planck~\cite{Collaboration:2015tu,Collaboration:2015tp}. Motivated by the functional form of the Surprise, we analyze how we can identify directions in parameter space that are causing it. We discuss how these results illustrate the versatility of the Surprise as an information metric for concordance in cosmology.

The paper is organized as follows. In section~\ref{sec:consistency} we review methods that were proposed for assessing the agreement of cosmological probes and compare them to the notion of Surprise. In section~\ref{sec:consistency_of_planck_and_wmap} we apply the Surprise to WMAP and Planck constraints. We conclude in section~\ref{sec:discussion}.

\section{Concordance of datasets}
\label{sec:consistency}

A number of approaches to quantifying the agreement between constraints from different datasets have been used in cosmology. Basically, each method consists of two steps. First, a measure for the agreement between constraints needs to be devised. Second, the measure needs to be interpreted on a reference scale that indicates the degree of (dis-)agreement between the datasets. In this section, we will revisit such a measure of concordance called Surprise that we introduced in~\cite{Seehars:2014ir}, before summarizing the main alternative measures proposed for cosmology.

\subsection{Cosmological parameter estimation}
\label{sub:parameter_estimation}

Inference in cosmology is typically based on the likelihood $p(\mathcal D|\Theta)$, the expected distribution of the data $\mathcal D$ given a model for the data that depends on parameters $\Theta$. The constraints on $\Theta$ from the data $\mathcal D$ can be derived via Bayes' theorem:
\begin{equation}
	p(\Theta|\mathcal D) = \frac{p(\mathcal D|\Theta)p(\Theta)}{p(\mathcal D)}
	\label{eq:bayes}
\end{equation}
where $p(\Theta)$ is the prior, $p(\Theta|\mathcal D)$ is the posterior, and $p(\mathcal D)$ is the evidence, defined as
\begin{equation}
	p(\mathcal D) = \int d\Theta p(\mathcal D|\Theta)p(\Theta).
\end{equation}
The evidence is the probability distribution for observing the data $\mathcal D$ given the prior knowledge on the parameters $\Theta$. 

It is instructive to study the details of the different measures for the agreement between constraints in the limit of a linear model and a Gaussian likelihood and prior. In this case, posteriors and evidences can be evaluated analytically based on a relation between the data covariance and parameter covariance given by the Fisher matrix. While the full multi-dimensional treatment is left for Appendix~\ref{sec:linear_gaussian_model}, we will illustrate some of the results with a simple, one-dimensional toy model. For this we assume that we collect two independent data points $d_1$ and $d_2$ which measure the same parameter $\theta$ and follow a Gaussian likelihood:
\begin{equation}
	\begin{aligned}
		p(d_i|\theta) &= \frac 1 {\sqrt{2\pi s_i^2}} \exp\left[-\frac 1 2\left(\frac {d_i - \theta} {s_i} \right)^2\right]\\
		&\equiv \mathcal N(d_i;\theta,s_i),
	\end{aligned}
\end{equation}
where $s_i$ is the standard deviation corresponding to the Gaussian error on the $d_i$ measurement. We assume that we start with some Gaussian prior distribution for $\theta$ with mean $\theta_p$ and variance $\sigma_p^2$. We also assume that $\sigma_p^2 \gg \sigma_i^2$, i.e. that the prior is weak compared to the measurements. In this case, the posterior is again Gaussian and given by
\begin{equation}
	p(\theta|d_i) \simeq \mathcal N\left(\theta;\theta_i,\sigma_i\right),
	\label{eq:1dpost}
\end{equation}
where $\theta_i = d_i$ and $\sigma_i = s_i$ for this simple toy model. Even though posterior and likelihood are identical in this case, we will keep distinguishing between $\theta_i$ and $d_i$ in order to make the difference between data and parameter space more apparent. 

\subsection{Surprise}
\label{sub:surprise}

The Surprise, introduced in~\cite{Seehars:2014ir}, is a measure for consistency between posterior distributions and hence operates in parameter space. We consider the case of two datasets $\mathcal D_1$ and $\mathcal D_2$ which are used to derive posteriors $p(\Theta|\mathcal D_1)$, $p(\Theta|\mathcal D_2)$, and eventually even the joint posterior $p(\Theta|\mathcal D_1, \mathcal D_2)$. The Surprise can be used to analyze either the compatibility of the separately analyzed posteriors $p(\Theta|\mathcal D_1)$ and $p(\Theta|\mathcal D_2)$, or the posteriors from a Bayesian update $p(\Theta|\mathcal D_1)$ and $p(\Theta|\mathcal D_1, \mathcal D_2)$. As the approach is equivalent for both cases, we will in the following denote both the individually and jointly analyzed posterior of $\mathcal D_2$ as $p(\Theta|\mathcal D_2)$ to simplify notation.

To measure the difference between $p(\Theta|\mathcal D_1)$ and $p(\Theta|\mathcal D_2)$ we use the relative entropy, or Kullback-Leibler divergence~\cite{Kullback:1951va}, $D(p(\Theta|\mathcal D_2)||p(\Theta|\mathcal D_1))$ defined as follows:
\begin{equation}
	D(p(\Theta|\mathcal D_2)||p(\Theta|\mathcal D_1)) = \int d\Theta\, p(\Theta|\mathcal D_2) \log \left(\frac{p(\Theta|\mathcal D_2)}{p(\Theta|\mathcal D_1)} \right).
\end{equation}
Here and in the following, $\log$ denotes the natural logarithm, so this is the definition for the relative entropy in units of \emph{nats}. To convert the relative entropy (and later the Surprise) into \emph{bits}, one can simply divide the values in nats by $\log(2)$.

The relative entropy is a convenient measure for comparing posteriors as it is zero if and only if both distributions are identical and is positive otherwise~\cite{Kullback:1951va}. It is however a directed measure, i.e. it is not invariant under interchanging $p(\Theta|\mathcal D_2)$ and $p(\Theta|\mathcal D_1)$. The relative entropy is furthermore independent of reparametrizations of the model~\cite{Kullback:1951va}. As the relative entropy measures both changes in precision and shifts in parameter space when updating or replacing data, it can be interpreted as the information gain when going from $p(\Theta|\mathcal D_1)$ to $p(\Theta|\mathcal D_2)$. It has been used in this sense to study the information gains from a historical sequence of CMB experiments~\cite{Seehars:2014ir} and from combining observations from a wide range of cosmological probes~\cite{Grandis:2015}.

To distinguish contributions to $D$ from gains in precision and shifts in parameter space, one can use constraints from one dataset to forecast the information gain that is expected for another dataset assuming that both are well described by the same model. Whenever the observed gains strongly differ from prior expectations, we interpret this as a systematic effect that is not explained by the model or included in the likelihood. From the posterior $p(\Theta|\mathcal D_1)$, and for consistent datasets, data $\mathcal D_2$ is anticipated to be a realization from
\begin{equation}
	p(\mathcal D_2|\mathcal D_1) = \int d\Theta\, p(\Theta|\mathcal D_1) p(\mathcal D_2 | \Theta)
\end{equation}
which is sometimes called the posterior predictive distribution of $\mathcal D_1$. On average, the posterior predictive $p(\mathcal D_2|\mathcal D_1)$ predicts a relative entropy of
\begin{equation}
	\ere \equiv \int d\mathcal D_2\, p(\mathcal D_2|\mathcal D_1) D(p(\Theta|\mathcal D_2)||p(\theta|\mathcal D_1)).
\end{equation}
We can hence compare the observed relative entropy $D$ to the expected relative entropy $\ere$. A measure of consistency, the Surprise of the constraints derived from $\mathcal D_2$, was consequently defined in~\cite{Seehars:2014ir} as:
\begin{equation}
	S \equiv D(p(\theta|\mathcal D_2)||p(\theta|\mathcal D_1)) - \ere.
\end{equation}
By construction, we expect $S$ to scatter around zero. If $S$ is positive, the posteriors are more different than expected a priori. If $S$ is negative, the constraints are more consistent than expected a priori.

The concept of expected relative entropy is also well known from Bayesian experimental design, where the aim is to design an experiment which maximizes $\ere$~\cite{Lindley1956}. Many algorithms have been proposed to perform this optimization in general non-linear cases (see e.g.~\cite{Drovandi2013320,Long201324,Huan2013288}). \citet{Huan2013288}, for example, rewrite the expected relative entropy for a joint analysis of $\mathcal D_1$ and $\mathcal D_2$ as:
\begin{equation}
	\begin{aligned}
		\ere = \int d\mathcal D_2\, \int d\Theta\, &\left(\log p(\mathcal D_2|\theta) - \log p(\mathcal D_2|\mathcal D_1) \right)\\ &\times p(\mathcal D_2|\theta) p(\Theta|\mathcal D_1).
	\end{aligned}
\end{equation}
Given a sample of size $n$ from $p(\Theta|\mathcal D_1)$, $\ere$ can then be estimated via
\begin{align}
	\ere &\approx \frac 1 n \sum_{i = 1}^n \left(\log p(\mathcal D_2^i|\Theta_i) - \log p(\mathcal D_2^i|\mathcal D_1)\right),\\
	p(\mathcal D_2^i|\mathcal D_1) &\approx \frac 1 n \sum_{j = 1}^n p(\mathcal D_2^i|\Theta_j),
\end{align}
where $\mathcal D_2^i$ is a sample from $p(\mathcal D_2^i|\Theta_i)$ for each $\Theta_i$ of the sample from $p(\Theta|\mathcal D_1)$. Together with approaches for calculating the relative entropy between non-Gaussian priors and posteriors (see e.g.~\cite{Grandis:2015, Skilling:ci}), this approach allows for efficient calculation of the Surprise in general, non-Gaussian cases. In order to interpret the observed Surprise, one would however also need to estimate the relative entropy for each data $\mathcal D_2^i$ to estimate the expected distribution of the Surprise.

In the standard model of a flat \lcdm cosmology, constraints on the parameters from the CMB are so stringent that an analysis assuming a linear model for the observables and Gaussianity of prior and likelihood is a good approximation. Any updates from or comparisons between CMB datasets can hence be approximated by this limit. Analytic results for the Surprise in this case have been shown to depend only on the covariances and means of the respective posteriors in~\cite{Seehars:2014ir}. For the Surprise, the result is simply given by
\begin{equation}
	S = \frac 1 2 \left(\Delta \mu \Sigma_1^{-1} \Delta \mu - \langle \Delta \mu \Sigma_1^{-1} \Delta \mu \rangle\right),
	\label{eq:surprise}
\end{equation}
where $\Delta \mu$ is the difference between the means of $p(\Theta|\mathcal D_1)$ and $p(\Theta|\mathcal D_2)$ and $\Sigma_1$ is the covariance of $p(\Theta|\mathcal D_1)$. $\langle \Delta \mu \Sigma_1^{-1} \Delta \mu \rangle$ is given by
\begin{equation}
	\langle \Delta \mu \Sigma_1^{-1} \Delta \mu \rangle = {\rm tr}\left(\mathbb{1} \pm \Sigma_2 \Sigma_1^{-1}\right),
\end{equation}
where $\Sigma_2$ is the covariance of $p(\Theta|\mathcal D_2)$ and the $-$ holds when $\mathcal D_2$ was used to update $p(\Theta|\mathcal D_1)$ and the $+$ when $p(\Theta|\mathcal D_2)$ is derived independently of $\mathcal D_1$. 

Equation~\eqref{eq:surprise} is a quite intuitive measure, as it measures the shift in the means relative to the covariance matrix of the prior information and is hence a natural generalization of the typical one-dimensional notion of $\Delta \mu / \sigma$ tensions. As shown in~\cite{Seehars:2014ir}, $S$ follows a generalized $\chi^2$ distribution and can be rewritten as a weighted sum of $\chi^2$ variables $Z_i$ with one degree of freedom,
\begin{equation}
	\begin{aligned}
		S &= \frac 1 2 \left(\sum_{i = 1}^d \lambda_i Z_i - {\rm tr}\left(\mathbb{1} \pm \Sigma_2 \Sigma_1^{-1}\right)\right) \\
		&= \frac 1 2 \sum_{i = 1}^d \lambda_i (Z_i - 1)
	\end{aligned}
	\label{eq:genchi2}
\end{equation}
where $\lambda_i$ are the eigenvalues of $\mathbb{1} \pm \Sigma_2 \Sigma_1^{-1}$ and $d$ is the dimensionality of the parameter space. 

Given a particular Surprise value $S$, one can use the cumulative distribution of the generalized $\chi^2$ random variable defined in equation \eqref{eq:genchi2} to calculate the probability for measuring a Surprise that deviates from zero by more than $S$. This probability is the so-called $p$-value for the hypothesis that the posteriors are derived from datasets which are consistent within the model. If the $p$-value is small, the datasets are unlikely to be consistent within the model. To compute the $p$-value for $S$ from mean and covariance of the posteriors, one can use an algorithm by~\citet{Davies:1980ty}, for instance, implemented in the \verb|R| package \verb|CompQuadForm| by~\citet{Duchesne:2010kq}.

The one-dimensional toy model of section~\ref{sub:parameter_estimation} is now a simple limit of equation~\eqref{eq:surprise}. In the case when $d_2$ is used to update $p(\theta|d_1)$, $S$ is given by
\begin{equation}
	S = \frac 1 2 \frac {\sigma_1^2} {\sigma_1^2 + \sigma_2^2} \left( \frac {(\theta_1 - \theta_2)^2} {\sigma_1^2 + \sigma_2^2} - 1\right).
	\label{eq:tmj}
\end{equation}
When $d_2$ is used to derive the posterior separately from $d_1$, the toy model results in:
\begin{equation}
	S = \frac 1 2 \left( \frac {(\theta_1 - \theta_2)^2} {\sigma_1^2} - \frac {\sigma_2^2 + \sigma_1^2} {\sigma_1^2} \right).
	\label{eq:tmr}
\end{equation}
In this toy model, the Surprise hence quantifies the difference between $d_1$ and $d_2$ by comparing their mean constraints $\theta_{1}$ and $\theta_{2}$ on the model. The error by which the difference is weighted depends on the scenario: deviations of $\theta_1$ and $\theta_2$ are compared to $\sigma_1^2 + \sigma_2^2$ for separately analyzed posteriors and to $\sigma_1^2$ only for joint analyses. The second summand of both expressions ensures that the Surprise is 0 on average.

As already mentioned earlier, the relative entropy, and hence the Surprise, are independent of the parametrization of the model. Indeed, equation~\eqref{eq:surprise} suggests a simple reparameterization of the parameter space when comparing constraints. An intuitive choice is a space $\Psi$ in which $p(\Theta|\mathcal D_1)$ has parameters that are uncorrelated, have unit variance, and mean zero. In this space, $\tilde \Sigma_1 = \mathbb 1$ by construction and the Surprise is given by
\begin{equation}
	S = \frac 1 2 \left(\tilde\mu_2^T \tilde \mu_2 - {\rm tr}\left(\mathbb 1 \pm \tilde\Sigma_2\right)\right),
	\label{eq:sstandardized}
\end{equation}
where $\sim$ indicates that the moments are now taken in the new space $\Psi$. Each shift away from zero in $\Psi$ independently contributes to $S$. As the constraints $p(\Psi|\mathcal D_1)$ are very simple, tensions between Gaussian constraints can then be simply visualized by shifts in the one-dimensional marginal distributions of the new parameters $\Psi$.

\subsection{Other measures}
\label{sub:measures_of_agreement}

The agreement between datasets within a given model can be estimated from the evidence in data space or the posterior in model space. In this section, we summarize two evidence and one posterior based method that have been applied to cosmology in the past. We illustrate the methods with the one-dimensional toy model introduced in section~\ref{sub:parameter_estimation}, but the results for the analysis of the general linear Gaussian model in higher dimensions are given in Appendix~\ref{sec:linear_gaussian_model}.

The most common tool~\cite{Marshall:2006hw,Amendola:2013ce,2014MNRAS.439.1855H,2014PhRvD..90f3501M,2015MNRAS.449.2405K,2015arXiv151000688R} for assessing the agreement between cosmological observations was introduced by~\citet{Marshall:2006hw}. It is defined as
\begin{equation}
	R \equiv \frac {p(\mathcal D_1, \mathcal D_2)} {p(\mathcal D_1) p(\mathcal D_2)},
\end{equation}
with the idea that it compares the evidence for $\mathcal D_1$ and $\mathcal D_2$ when both have to be described by the same parameters of the model (numerator) to the evidence when each dataset is allowed to be described by different parameter values (denominator). The value $R$ is then interpreted on Jeffrey's scale, indicating consistent datasets when $R > 1$ and inconsistencies otherwise. It can be used to compare multi-dimensional, non-Gaussian constraints and numerical techniques for estimating the Bayesian evidence are well established.

For the one-dimensional toy model introduced in section~\ref{sub:parameter_estimation}, $R$ evaluates to the following relation~\footnote{The asymmetry between $d_1$ and $d_2$ in this expression is an artifact of the limit of a wide prior. As the expressions are much simpler in this limit we decided to anyways show these results. The general expression in Appendix~\ref{sec:linear_gaussian_model} does not show this asymmetry and all conclusions are not affected by the limit.}:
\begin{equation}
	\begin{aligned}
		\log R = - &\frac 1 2 \left[\log \left( \frac{s_2^2 + \sigma_1^2} {\sigma_p^2} \right) +\right.\\
		&+ \left. \frac {\left(d_2 - \theta_1\right)^2} {s_2^2+\sigma_1^2} - \frac {(d_2 - \theta_p)^2} {\sigma_p^2}\right]
	\end{aligned}
	\label{eq:gr}
\end{equation}
The terms in the bottom row of equation \eqref{eq:gr} compare $d_2$ to the mean model $\theta_1$ from the posterior of $d_1$ and to the mean model $\theta_p$ of the prior. For consistent datasets, i.e. when $d_1$ and $d_2$ are drawn from $p(d_1, d_2)$, we expect that those cancel each other on average. The term in the top row, however, is data independent and affected only by changes in the variance. For this example, $\log  R$ is therefore expected to vary about the variance dependent term. Hence, the weaker the prior, the more one tends to underestimate the tensions when interpreting $\log R$ relative to $1$. The example shows that interpreting the ratio independent of specific properties of prior and posterior can be non-trivial, as deviations from $ R = 1$ can be attributed to inconsistencies between the data and differences between variances at the same time. 

To analyze the agreement between constraints on the Hubble constant and the age of the Universe from low-redshift measurements and CMB observations, \citet{Verde:2013hp} use a modified version of the $R$ measure. The idea is to perform a translation in the parameters $\Theta$ of one of the likelihoods $p(\mathcal D_i|\Theta)$, with $\bar p(\mathcal D_i|\Theta)$ being the shifted distribution. For the translation where the maxima of the two distributions coincide, the evidence is written as
\begin{equation}
	\left.\bar p(\mathcal D_1, \mathcal D_2)\right|_{{\rm max}\, \mathcal D_1  = {\rm max}\, \mathcal D_2} = \int d\Theta\, \bar p(\mathcal D_2|\Theta) p(\Theta|\mathcal D_1),
\end{equation}
where we chose to shift $\bar p(\mathcal D_2|\Theta)$ in this example. The measure of tension by \citet{Verde:2013hp} is then defined as
\begin{equation}
	T \equiv \frac {\left.\bar p(\mathcal D_1,\mathcal D_2)\right|_{{\rm max}\, \mathcal D_1  = {\rm max}\, \mathcal D_2}} {p(\mathcal D_1, \mathcal D_2)}.
\end{equation}
The intuition is that $\left.\bar p(\mathcal D_1,\mathcal D_2)\right|_{{\rm max}\, \mathcal D_1  = {\rm max}\, \mathcal D_2}$ is the evidence for maximally consistent datasets $\mathcal D_1$ and $\mathcal D_2$, to which the actual evidence is compared. $T$ is interpreted on a slightly modified Jeffrey's scale indicating tensions when $\log T > 1$.

Looking at the toy model, we find for $\log T$
\begin{equation}
	\log T = \frac 1 2 \frac {\left(d_2 - \theta_1\right)^2} {\sigma_1^2+\sigma_2^2}.
\end{equation}
Similarly to $R$, this measure compares the deviations between $d_2$ and the mean model for the data from the posterior of $d_1$. It however does neither depend on the prior nor show the data independent variance term, implying that $\log T$ is fairly problem independent for this toy model. On average, we expect this expression to be of order 1, or of order $d$ for general, $d$-dimensional parameter spaces, for consistent data. We hence find that $\log T$ tends to larger values as the size of the parameter space increases. Interpreting it relative to $1$ will lead to an overestimation of tensions for parameter spaces with dimensionality greater than 2.

\citet{MacCrann:2015ke} define a measure of consistency between two datasets in parameter space by looking at the best-fit point $\Theta_{\rm joint}$ of the joint constraints and its likelihood $p(\mathcal D_i|\Theta_{\rm joint})$ given either $\mathcal D_1$ or $\mathcal D_2$. Within a sample $\{\Theta_i\}_{i = 1}^N$ from the posterior $p(\Theta|\mathcal D_i)$ one can compute the percentile of likelihoods $p(\mathcal D_i|\Theta_i)$ that are smaller than $p(\mathcal D_i|\Theta_{\rm joint})$. This percentile is then interpreted as the confidence for the two datasets being consistent. This technique is able to compare multi-dimensional, non-Gaussian constraints. The $p$-value of the best-fit point can be estimated from Monte Carlo Markov chains (MCMCs) of the individual posteriors.

For the toy model, the best fit point of the joint distribution is simply the mean $\theta_{\rm joint}$ of the joint posterior. Without loss of generality, we look at the expected distribution of $\theta_{\rm joint}$ as predicted by the posterior $p(\theta|d_1)$ of $d_1$ and the likelihood $p(d_2|\theta)$ assuming that the model correctly describes both sets of data. We find that, for consistent data, $\theta_{\rm joint}$ is expected to be drawn from the following normal distribution:
\begin{equation}
	\theta_{\rm joint} \leftarrow \mathcal N\left(\theta;\theta_1,\sigma_1\sqrt{\frac{\sigma_1^2}{\sigma_1^2+\sigma_2^2}}\right).
	\label{eq:ptj}
\end{equation}
Comparing the distribution for $\theta_{\rm joint}$~\eqref{eq:ptj} to the posterior $p(\theta|d_1)$ of $d_1$~\eqref{eq:1dpost}, we find that even though the means coincide, the expected standard deviation of $\theta_{\rm joint}$ is smaller than $\sigma_1$. We hence expect that, on average, the likelihood $p(d_1|\theta_{\rm joint})$ of the best fit point is larger than the likelihood of a sample from the posterior $p(\theta|d_1)$. The percentile of $p(d_1|\theta_{\rm joint})$ will therefore underestimate the tension introduced by the update. Only in the limit where $\sigma_1^2 \gg \sigma_2^2$, i.e. where the constraints from $d_1$ are weak compared to the joint constraints, is $\theta_{\rm joint}$ expected to be distributed approximately according to $p(\theta|d_1)$.

\subsection{Comparison of measures}
\label{sub:comparison}

In this short summary, we have seen that it is possible to define and estimate a variety of measures of agreement between general, non-Gaussian constraints. Interpreting them on a fixed, problem-independent scale can however over- or underestimate the tensions in the data. 

While computationally expensive, a potential way out are simulation based approaches, where the measure is calibrated from Monte Carlo realizations of mock data from the posterior predictive of one of the constraints in question. Following~\citet{Huan2013288}, such an approach was briefly outlined in~\ref{sub:surprise} for the Surprise. Also~\citet{Larson:2015gk}, for example, take such an approach to comparing the constraints from WMAP~9 and Planck~13 and compare the maximum likelihood estimates of each experiment based on a simulation study. The authors create Monte Carlo realizations of CMB maps from an input power spectrum and process these maps into WMAP- and Planck-like observations by adding the corresponding noise features and masks. They then estimate the spectra of the simulated maps and find the best-fit points for each simulated WMAP and Planck map. The sample of differences between the simulated best-fit points is then compared to the observed best-fit difference on a parameter by parameter basis. A further method for dealing with potentially inconsistent datasets, proposed by~\citet{2002MNRAS.335..377H}, is to allow for additional weight parameters when combining the likelihoods in the estimation procedure. In this approach, datasets that introduce tensions in the parameter constraints are down-weighted in order to avoid biases from systematics.

Another possibility is to focus on situations where analytic results from a linear Gaussian model are a good approximation. This limit tends to be a good description for flat \lcdm constraints as soon as CMB data from Planck or WMAP is involved. In these cases, the distributions and expectations for most of the reviewed measures can be derived (see Appendix~\ref{sec:linear_gaussian_model}). The strength of the Surprise in this scenario is its flexibility: it can be both used to analyze separately derived constraints and constraints from a Bayesian update. Operating in parameter space, it furthermore only depends on mean and covariance of prior and posterior which can be robustly estimated from standard MCMC samples. Finally, the Surprise follows a relatively simple distribution given by the weighted sum of $d$ $\chi^2$ distributed variables, where $d$ is the dimension of the parameter space. Given again only the covariance of prior and posterior, the $p$-value of the observed Surprise can be evaluated with standard numerical tools~\cite{Davies:1980ty, Duchesne:2010kq}. We provide a \verb|Python| module for calculating the relative entropy, the Surprise and its $p$-value in the linear Gaussian case at \url{https://github.com/seeh/surprise}.

\section{Application to Planck and WMAP}
\label{sec:consistency_of_planck_and_wmap}

The first application of the Surprise in~\cite{Seehars:2014ir} was a study of the agreement between constraints on a flat \lcdm cosmology from a historical sequence of CMB experiments. One of the main results of this study was the detection of a significant Surprise when comparing constraints from WMAP~9 data~\cite{Bennett:2012wy, Hinshaw:2013dd} and the constraints from the 2013 release of Planck data~\cite{PlanckCollaboration:2014km, PlanckCollaboration:2014kt}. 

In this section, we revisit this comparison in light of the 2015 Planck release and perform a more detailed study of the Surprise between WMAP and Planck. We analyze the Surprise between the posteriors derived from the WMAP~9 data on temperature and polarization~\cite{Bennett:2012wy, Hinshaw:2013dd}, the Planck~13 temperature data~\cite{PlanckCollaboration:2014km, PlanckCollaboration:2014kt} together with low-$\ell$ WMAP polarization, and finally the Planck~15 data on temperature and polarization~\cite{Collaboration:2015tu}. We work in the six-dimensional flat \lcdm parameter space given by the Hubble constant $H_0$, the cold dark matter density today $\Omega_ch^2$, the baryonic matter density $\Omega_bh^2$, the optical depth to reionization $\tau$, and the amplitude $A_s$ and spectral index $n_s$ of the power spectrum of primordial curvature fluctuations. As highlighted in section~\ref{sub:surprise}, we need mean and covariance of these parameters for each of the posteriors. We estimate those moments from Monte Carlo Markov chains that were provided by the Planck team for the 2013 and 2015 Planck releases~\cite{PlanckCollaboration:2014dt, Collaboration:2015tp}. 

Due to cosmic variance correlations between the data from WMAP and Planck, the joint likelihood of both datasets cannot be factorized into the individual likelihoods. For a joint analysis, one would rather have to find the correct joint likelihood function that takes the cosmic variance correlations into account. For simplicity, we hence consider the posteriors from the separately analyzed data sets. When estimating expected relative entropy and Surprise, however, we have to ignore the correlations due to cosmic variance and the large scale WMAP polarization data used in the Planck 2013 analysis to constrain $\tau$. If we were able to take the correlations into account, they would decrease the expected relative entropy and increase the Surprise. 

\begin{figure}[t]
	\centering
	\includegraphics[width=1.\linewidth]{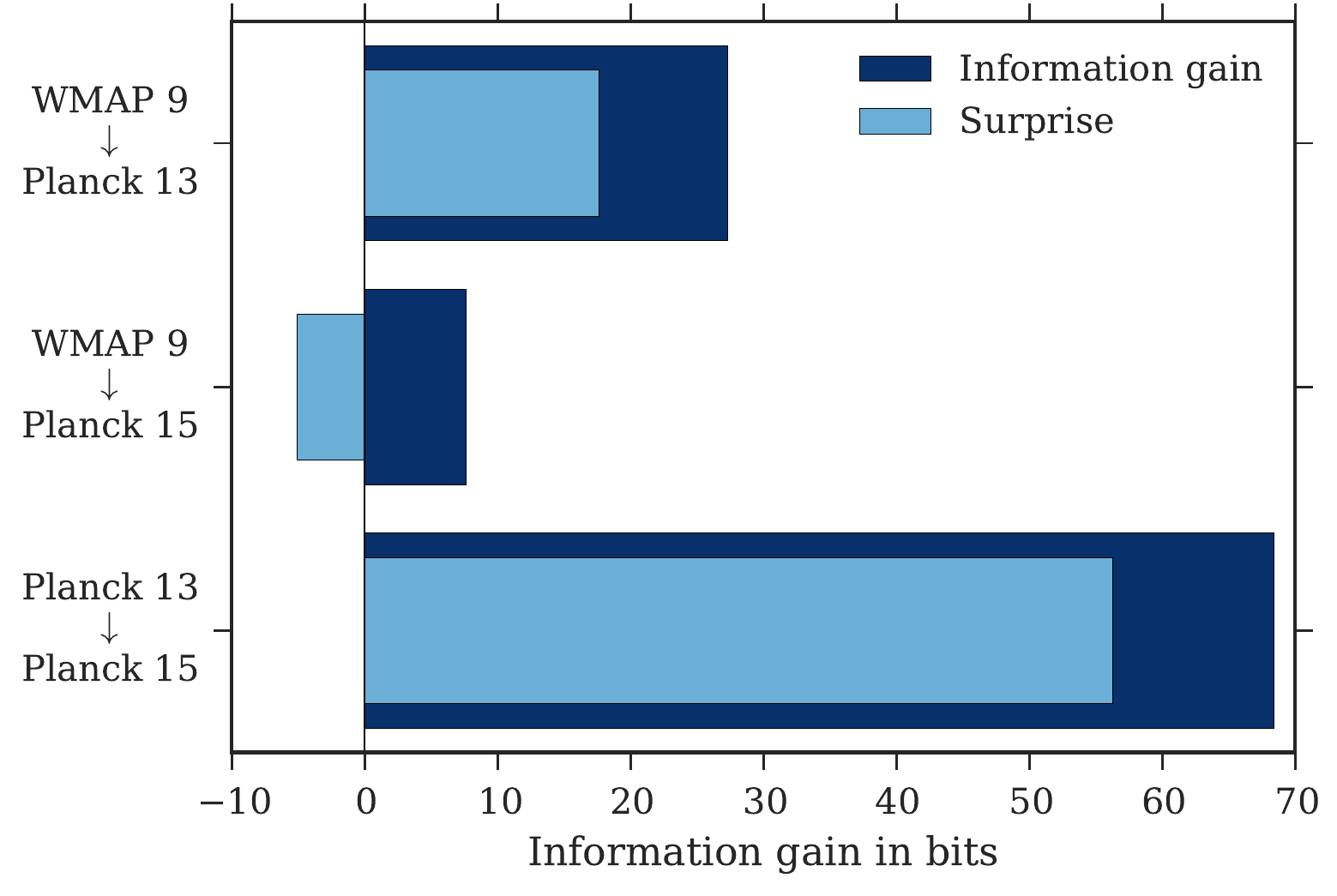}
	\caption{Information gains and Surprise values for different combinations of WMAP and Planck results. The dark blue bars show the overall information gain, while the light blue bars show the Surprise. We can see that, while Planck~13 is in strong tension with both WMAP~9 and Planck~15, the constraints from WMAP~9 and Planck~15 agree very well.}
	\label{fig:barplot}
\end{figure}

\begin{table}[t]
\centering
\caption{Relative entropy and Surprise estimates in bits for comparisons between WMAP and Planck data. The $p$-value is the prior probability of observing a Surprise that is greater or equal (less or equal) than $S$ if $S$ is greater (smaller) than zero when assuming consistent data.}
\begin{ruledtabular}
\begin{tabular}{rclrrrl}
	\multicolumn{3}{c}{Data combination} & \multicolumn{1}{c}{$D$} & \multicolumn{1}{c}{$\langle D \rangle$} & \multicolumn{1}{c}{$S$} & $p$-value\\\hline
	WMAP 9 &$\rightarrow$& Planck 13 & $27.3$ & $9.6$ & $17.6$ & $2 \times 10^{-3}$\\
	WMAP 9 &$\rightarrow$& Planck 15 & $7.6$ & $12.6$ & $-5.1$ & $7 \times 10^{-2}$\\
	Planck 13 &$\rightarrow$& Planck 15 & $68.4$ & $12.0$ & $56.3$ & $4 \times 10^{-5}$\\
\end{tabular}
\end{ruledtabular}
\label{tab:relent}
\end{table}

Table~\ref{tab:relent} shows the numerical values for the observed relative entropy, expected relative entropy, Surprise, and $p$-value of the Surprise. The values are visualized in Figure~\ref{fig:barplot}. As already found in~\cite{Seehars:2014ir}, the large Surprise ($S = 17.6$) when going from WMAP~9 to Planck~13 constraints indicates significant tension between those datasets ($p$-value of $0.002$)~\footnote{Note that the numbers of the comparison between WMAP~9 and Planck~13 do not perfectly agree with the numbers quoted in~\cite{Seehars:2014ir} because the official Planck chains use only low-$\ell$ WMAP polarization while~\cite{Seehars:2014ir} used the Planck temperature data together with the full WMAP polarization data.}. In light of the 2015 release of Planck, however, the large Surprise between WMAP~9 and Planck~13 has vanished, implying that the constraints of WMAP~9 and Planck~15 are in good agreement ($S = -5.1$). The Surprise between the Planck~13 and Planck~15 releases ($S = 56.3$) furthermore shows that the shifts in parameter space are significantly larger than expected a priori ($p$-value of $4 \times 10^{-5}$).

\begin{figure*}[t]
	\centering
	\includegraphics[width=.62\linewidth]{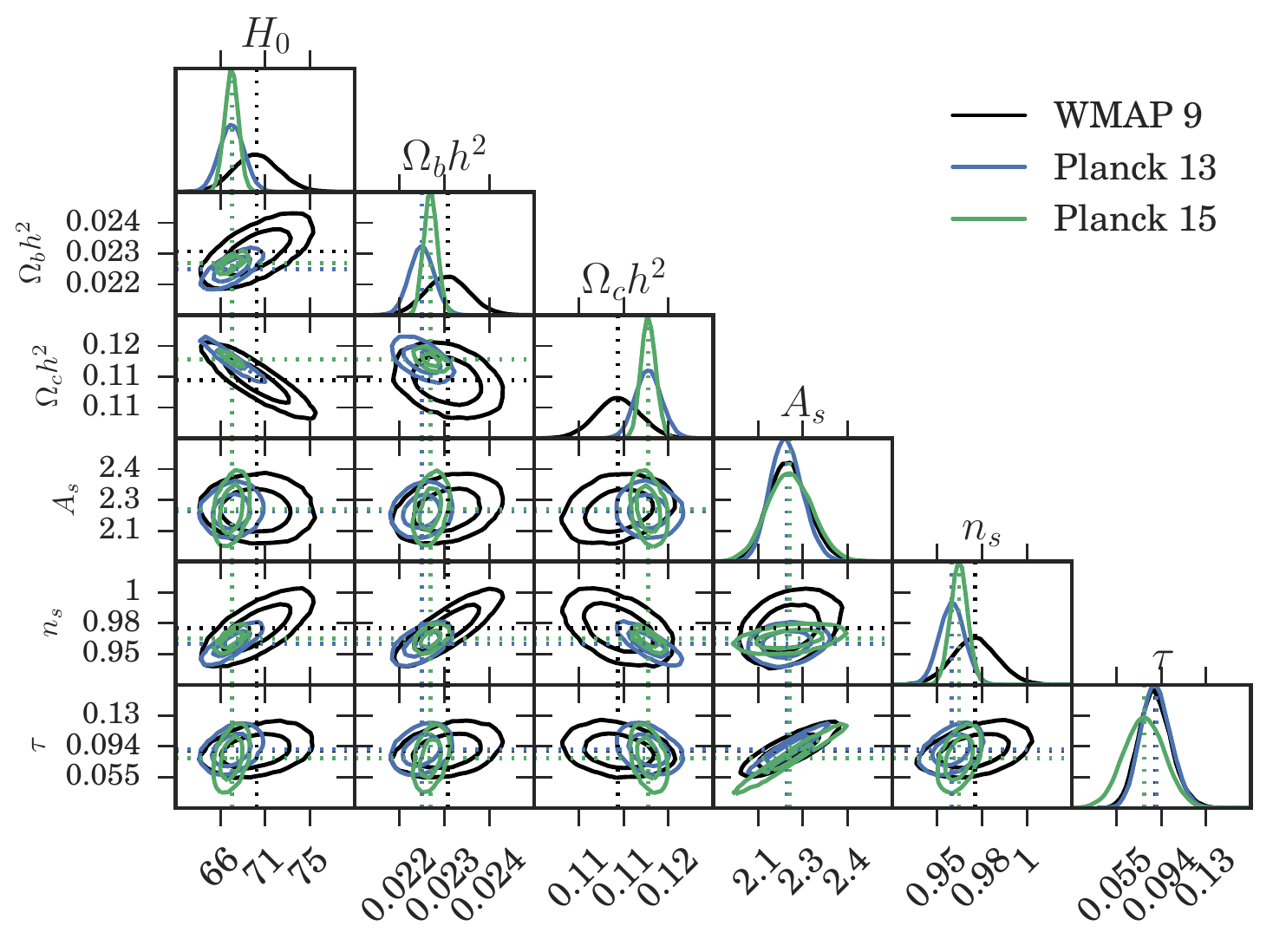}
	\caption{Marginalized posteriors of WMAP~9 (black), Planck~13 (blue), and Planck~15 (green) constraints in the parameters of a flat \lcdm~model. The contours cover 68\% and 95\% of the overall posterior volume and the dotted lines show the means.}
	\label{fig:ocontours}
\end{figure*}

\begin{figure}[t]
	\centering
	\includegraphics[width=1.\linewidth]{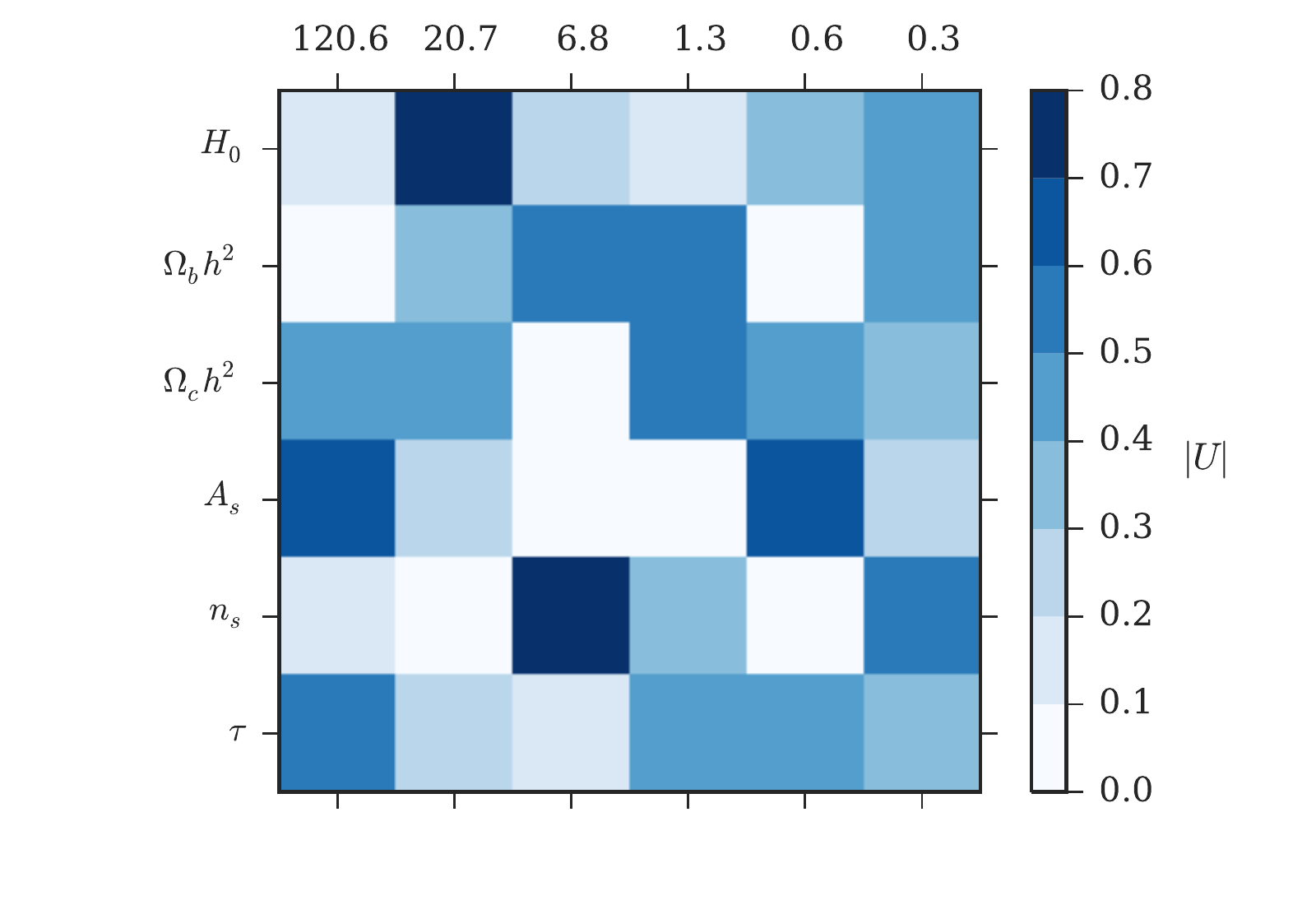}
	\caption{Matrix plot of $U$ (defined in equation \eqref{eq:defU}), showing the eigenvectors (columns) of the correlation matrix of the WMAP~9 posterior in the cosmological parameter space together with the inverse eigenvalues given by the diagonal of $P$ (column labels). We plot the absolute value of $U$ to show the strength with which each cosmological parameter contributes to the eigenvectors that define the new parameter space (see color scale). For the first new parameter (labeled by 120.6), for example, we can see that it gets the strongest contribution from $A_s$, but furthermore points into the $\tau$ and $\Omega_c h^2$ direction.}
	\label{fig:components}
\end{figure}

Looking at the marginalized constraints in Figure~\ref{fig:ocontours}, however, these results are hard to relate to the seemingly mild changes between the Planck~13 and Planck~15 results. To better understand the Surprise between the Planck~13 constraints and the WMAP~9 and Planck~15 results, we argued in section~\ref{sub:surprise} that the functional form of the Surprise suggests to standardize the WMAP constraints. Standardization means that we reparametrize our theory such that the WMAP~9 constraints in the new parameters are uncorrelated with mean 0 and standard deviation 1. The advantage of the standardized parameter is that each component independently contributes to the Surprise (see equation \eqref{eq:sstandardized}), potentially allowing us to identify directions in parameter space that generate the large Surprise.

Standardization is not unique and we choose to work in the eigenbasis of the correlation matrix for numerical stability. The correlation matrix is related to the covariance matrix via ${\rm Corr}(\theta_i, \theta_j) = {\rm Cov}(\theta_i, \theta_j)/\sigma(\theta_i) \sigma(\theta_j) $, where $\sigma(\theta_i)$ is the standard deviation of $\theta_i$. Given the eigendecomposition of the correlation matrix ${\rm Corr}$ of the parameters $\Theta$:
\begin{equation}
	{\rm Corr} = U^T P U
	\label{eq:defU}
\end{equation}
and a diagonal matrix $F$ containing the standard deviations of $\Theta$, the following parameters $\Psi$ follow a standard normal WMAP~9 posterior:
\begin{equation}
	\Psi = \left(\sqrt{P}\right)^{-1}UF^{-1}\left(\Theta - \mu(\Theta)\right).
	\label{eq:eb}
\end{equation}
Here, $\sqrt{P}$ is simply containing the square roots of the eigenvalues $P$ and $\mu(\Theta)$ contains the means of the parameters $\Theta$. The matrix $U$ defines how the new parameters $\Psi$ are related to the original cosmological parameters. It is shown in Figure~\ref{fig:components} together with the inverse eigenvalues $P^{-1}$ for each eigenvector. We use the inverse eigenvalues in the following to label the new parameters.

\begin{figure*}[t]
	\centering
	\includegraphics[width=.62\linewidth]{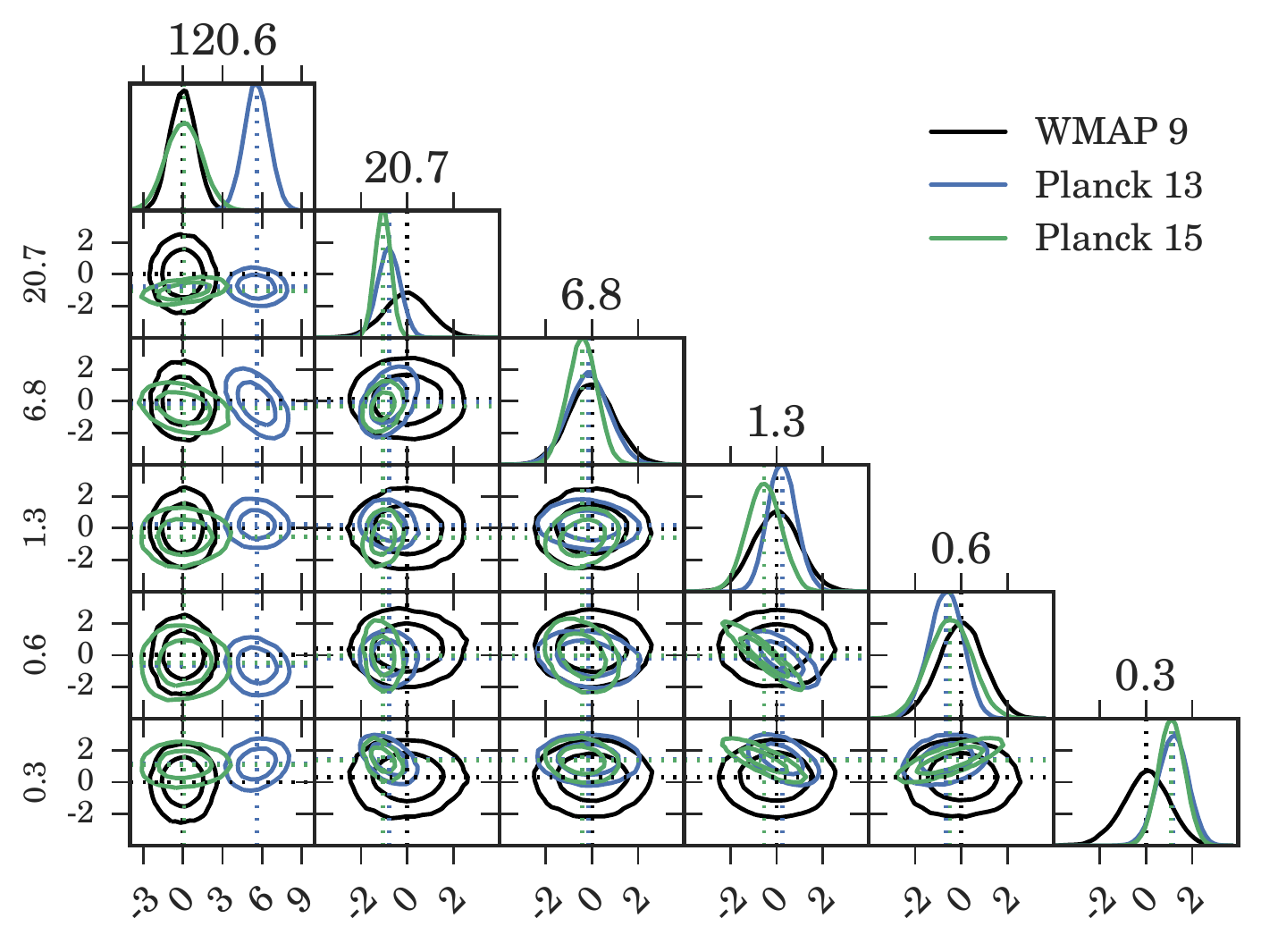}
	\caption{Marginalized posteriors of WMAP~9 (black), Planck~13 (blue), and Planck~15 (green) constraints in the parametrization $\Psi$ given by equation~\eqref{eq:eb}. The contours cover 68\% and 95\% of the overall posterior volume and the dotted lines show the means. The labels show the inverse eigenvalues of the correlation matrix given by the diagonal of $P$ defined in equation \eqref{eq:defU}. The relation between $\Psi$ and the cosmological parameters is shown in Figure \ref{fig:components}. By construction, the WMAP~9 constraints on $\Psi$ are uncorrelated, have mean 0 and standard deviation 1. Each deviation from $0$ in the means of the Planck constraints independently contributes to the Surprise between WMAP and Planck.}
	\label{fig:contours}
\end{figure*}

The marginalized posteriors in the new parameter space $\Psi$ are shown in Figure~\ref{fig:contours}. Note that Figure~\ref{fig:contours} shows the same constraints as Figure~\ref{fig:ocontours} in a different parametrization of the \lcdm model. Note also that the new parametrization is entirely derived from the WMAP~9 constraints without any knowledge of the Planck data. By construction, the WMAP~9 constraints have mean $0$, standard deviation $1$, and show no correlations between the individual $\psi_i$ components of $\Psi$. The striking feature of Figure~\ref{fig:contours} is the deviation of the Planck~13 constraints from the WMAP~9 mean by more than $5 \sigma$ in the direction of the parameter space with the dominant eigenvalue. The Planck~15 constraints, on the other hand, agree with the WMAP~9 constraints in this direction to great accuracy, thereby explaining the vanishing Surprise.

Looking at the eigenvector that corresponds to the largest eigenvalue (see first column of Figure~\ref{fig:components}), we see that the direction that is responsible for the tension between Planck~13 and the WMAP~9 and Planck~15 constraints is dominated by $A_s$ with additional contributions from $\tau$ and $\Omega_ch^2$. \citet{Larson:2015gk} come to a very similar conclusion based on the simulation-based approach to comparing WMAP and Planck data that was outlined in section~\ref{sub:comparison}. They point out that there is a mismatch in amplitude between Planck~13 data and WMAP~9 for small $\ell$ which dominate the constraints on the cosmological parameters and also conclude that the shifts in the $\Omega_ch^2$-$A_s$ direction are larger than expected when taking cosmic variance correlations into account. $A_s$ and $\tau$ dominantly affect the overall amplitude of the temperature power spectrum. The Planck team indeed changed their calibration scheme from the 2013 to the 2015 release~\cite{Collaboration:2015ta,Collaboration:2015wu}, and the strong change in the Surprise is most likely a direct consequence of this change in the data. 

It is not easy to interpret the actual magnitude of the negative Surprise between WMAP~9 and Planck~15. In principle, a negative Surprise indicates that the agreement between the constraints is better than expected from statistical fluctuations. The $p$-value of $0.07$ for the Surprise being $-5.1$ or smaller is estimated assuming independence of Planck and WMAP power spectra. However, the measurements are correlated due to cosmic variance as they both measure the same CMB. We would therefore expect the measurements to be more consistent than predicted from the WMAP~9 posterior. The confidence of the over-consistency between WMAP~9 and Planck~15 is small anyways and taking the correlations into account would most likely diminish it.

\section{Conclusions}
\label{sec:discussion}

In this work, we have revisited the Surprise~\cite{Seehars:2014ir}, a measure for the agreement between cosmological datasets within a cosmological model. The Surprise is based on the relative entropy between two posteriors and is a global measure of consistency over the entire parameter space. In the past, methods based on evidence ratios~\cite{Marshall:2006hw,Amendola:2013ce,2014MNRAS.439.1855H,2014PhRvD..90f3501M,2015MNRAS.449.2405K,2015arXiv151000688R,Verde:2013hp} or the likelihood of the joint best-fit to two datasets within the posteriors of the individual constraints~\cite{MacCrann:2015ke} have been used to compare different sets of cosmological data. For general non-Gaussian distributions, these measures can be more straightforward to calculate than the Surprise. However, we conclude from a Gaussian toy model analysis that interpreting those measures on a fixed, problem independent scale can be misleading. To make precise statements on the compatibility of datasets for arbitrary constraints, a calibration step based on Monte Carlo simulations would be needed.

Alternatively, one can focus on situations where this calibration can be done analytically. CMB observations, for example, are constraining flat \lcdm to a precision where posteriors are well approximated by Gaussian distributions. Whenever updating or comparing CMB constraints on \lcdm, analytical results for Gaussian distributions can hence be used to simplify the interpretation. The Surprise is particularly useful in this limit. Being derived from a general framework that measures both tensions and gains in precision between posterior distributions in the same units of bits, it can handle Bayesian updates as well as independently analyzed data. It furthermore depends only on the mean and covariance of the posteriors which can be robustly estimated from standard MCMC samples. For consistent data, the Surprise is expected to follow a relatively simple generalized $\chi^2$ distribution. The $p$-value for observing a given Surprise, assuming that the data is consistent within the model, can be estimated from the posterior moments. For convenience, we provide a \verb|Python| module for calculating the relative entropy, the Surprise and its $p$-value in the linear Gaussian case at \url{https://github.com/seeh/surprise}.

The Surprise was first applied to a historical sequence of CMB constraints in~\cite{Seehars:2014ir}, where a significant disagreement between WMAP~9 and the constraints from the 2013 release of Planck data was detected ($S = 17.6$ bits, $p$-value of $0.002$). When taking the 2015 release of Planck data into account, we find that Planck~2015 deviates strongly from the Planck~2013 results ($S = 56.3$ bits), but is in good agreement with the WMAP~9 constraints as indicated by a negative Surprise ($S = -5.1$ bits). By analyzing the posteriors in the principal components of the WMAP~9 constraints, we were able to detect the direction in parameter space that is the primary cause for the Surprise in the update from WMAP~9 to Planck~13. It is mainly composed by $A_s$, $\tau$, and $\Omega_c h^2$ and shifts along this direction lead to a change in the amplitude of the predicted temperature power spectrum. The inconsistency between Planck~13 and WMAP~9 as detected by the Surprise is hence most likely the consequence of a systematic in the calibration of the Planck data which was resolved in the 2015 release of the Planck team~\cite{Collaboration:2015ta,Collaboration:2015wu,Larson:2015gk}.

Our study shows that the Surprise is a reliable measure of agreement between cosmological constraints. While there exist other measures for the agreement between datasets that can be easier to calculate for general distributions, the strength of the Surprise is its interpretability for well constrained models. As more cosmological probes are able to put tight constraints on \lcdm parameters, estimating the Surprise between the constraints can help to detect systematic issues in data or model in the future.

\begin{acknowledgments}
We acknowledge use of the Planck Legacy Archive. Planck (\url{http://www.esa.int/Planck}) is an ESA science mission with instruments and contributions directly funded by ESA Member States, NASA, and Canada. This work was in part supported by the Swiss National Science Foundation (grant number $200021\_143906$).
\end{acknowledgments}

\bibliography{references}

\appendix

\section{Linear Gaussian model}
\label{sec:linear_gaussian_model}

Formally, we consider two experiments that are described by data $\mathcal D_i$ and likelihood $p(\mathcal D_i|\Theta)$ where $\Theta$ are the parameters of a model for the data and $i = 1,2$. Using Bayes' theorem, both experiments can put individual constraints on the parameters of the model via
\begin{equation}
	p_i \equiv p(\Theta|\mathcal D_i) = \frac {p(\mathcal D_i|\Theta)p(\Theta)} {p(\mathcal D_i)},
\end{equation}
where we call $p$, $p_i$, and $p(\mathcal D_i)$ prior, posterior, and evidence, respectively. The evidence is the prior probability for observing data $\mathcal D_i$ given by:
\begin{equation}
	p(\mathcal D_i) = \int d\Theta\, p(\mathcal D_i|\Theta)p(\Theta).
	\label{eq:evidence}
\end{equation}

Both sets of data can also be used to put joint constraints on the model parameters. This either requires knowledge of the joint likelihood $\mathcal L(\Theta;\mathcal D_1,\mathcal D_2) = p(\mathcal D_1,\mathcal D_2|\Theta)$ or independence of the measurement in the sense that the joint likelihood can be factorized $\mathcal L(\Theta;\mathcal D_1,\mathcal D_2) = \mathcal L_1(\Theta;\mathcal D_1)\mathcal L_2(\Theta;\mathcal D_2)$. If independence is assumed, the joint posterior is given by:
\begin{equation}
	p_{\rm joint}(\Theta|\mathcal D_1,\mathcal D_2) = \frac {\mathcal L_1(\Theta;\mathcal D_1)\mathcal L_2(\Theta;\mathcal D_2)p(\Theta)} {p(\mathcal D_1,\mathcal D_2)}.
\end{equation}
It is worth noting that this is formally equivalent to using the constraints from one set of data as a prior for the analysis of the other. For the evidence, a similar statement holds:
\begin{equation}
	\begin{aligned}
		p(\mathcal D_1,\mathcal D_2) &= \int d\Theta\, p(\mathcal D_1|\Theta)p(\mathcal D_2|\Theta)p(\Theta)\\ 
		&= p(\mathcal D_1) \int d\Theta\, p(\mathcal D_2|\Theta)p(\Theta|\mathcal D_1) \\
		&\equiv p(\mathcal D_1)p(\mathcal D_2|\mathcal D_1).
	\end{aligned}
	\label{eq:evs}
\end{equation}

For the case of a Gaussian prior with mean $\Theta_p$ and covariance $\Sigma_p$, a linear model for the data $F_i(\Theta) = F_i^0 + M_i\Theta$, and a Gaussian likelihood with covariance $C_i$, the posterior is Gaussian with moments~\cite{Seehars:2014ir}:
\begin{align}
	\Sigma_i &= \left(\Sigma_p^{-1} + M_i^TC_i^{-1}M_i\right)^{-1},\\
	\Theta_i &= \Theta_p + \Sigma_i M_i^T C_i^{-1}\left(\mathcal D_i - F(\Theta_p)\right).
\end{align}
Also $p(\mathcal D_i)$ is Gaussian (in the data) with mean $F(\Theta_p)$ and covariance $C_i + M_i\Sigma_pM_i^T$ and we use the following notation for this:
\begin{equation}
	p(\mathcal D_i) = \mathcal N(\mathcal D_i;F(\Theta_p),C_i + M_i\Sigma_pM_i^T)
\end{equation}
We now have all the ingredients to study the various techniques outlined in section~\ref{sub:measures_of_agreement} in more detail.

We start with the $R$ measure by~\citet{Marshall:2006hw}, defined as
\begin{equation}
	R = \frac {p(\mathcal D_1, \mathcal D_2)} {p(\mathcal D_1)p(\mathcal D_2)}.
\end{equation}
Using equation~\eqref{eq:evs}, we can rewrite $R$ as
\begin{equation}
	R = \frac {p(\mathcal D_2|\mathcal D_1)} {p(\mathcal D_2)},
\end{equation}
i.e. as the ratio between the evidence for $\mathcal D_2$ given the posterior from $\mathcal D_1$ to the evidence for $\mathcal D_2$ from the prior. In the linear Gaussian model, we know both distributions:
\begin{align}
	p(\mathcal D_2|\mathcal D_1) &= \mathcal N(\mathcal D_2; F_2(\Theta_1), C_2 + M_2\Sigma_1M_2^T)\\
	p(\mathcal D_2) &= \mathcal N(\mathcal D_2; F_2(\Theta_p), C_2 + M_2\Sigma_pM_2^T)
\end{align}
Using this, we can rewrite the logarithm of $R$ as
\begin{equation}
	\begin{aligned}
		&\log R = -\frac 1 2 \left[\log \frac {\det (C_2 + M_2\Sigma_1M_2^T)} {\det (C_2 + M_2\Sigma_pM_2^T)}\right. \\
		&+ (\mathcal D_2 - F_2(\Theta_1))^T (C_2 + M_2\Sigma_1M_2^T)^{-1}(\mathcal D_2 - F_2(\Theta_1))\\
		&\left.- (\mathcal D_2 - F_2(\Theta_p))^T(C_2 + M_2\Sigma_pM_2^T)^{-1}(\mathcal D_2 - F_2(\Theta_p))\right].
	\end{aligned}
	\label{eq:logr}
\end{equation}
Equation~\eqref{eq:logr} hence depends on the difference between the covariances of $p(\mathcal D_2|\mathcal D_1)$ and $p(\mathcal D_2)$ as well as on the difference between the model of prior and posterior means $\Theta_q$ and $\Theta_1$ and the data $\mathcal D_2$. Note in particular that $\log R$ depends on the prior moments $\Theta_p$ and $\Sigma_p$ because of $p(\mathcal D_2)$. This is a well known problem of the evidence when used with a wide prior that is supposed to be uninformative. On average, i.e. when calculating the expected value of \eqref{eq:logr} under $p(\mathcal D_1, \mathcal D_2)$, all the data dependent terms cancel each other and we find
\begin{equation}
	\begin{aligned}
	\langle \log R \rangle &\equiv \int d\mathcal D_1\,d\mathcal D_2\,p(\mathcal D_1,\mathcal D_2) \log R \\
	&= -\frac 1 2 \log \frac {\det (C_2 + M_2\Sigma_1M_2^T)} {\det (C_2 + M_2\Sigma_pM_2^T)}
	\end{aligned}
\end{equation}

The second evidence-based measure by~\citet{Verde:2013hp} is defined as:
\begin{equation}
	T = \frac {\left.\bar p(\mathcal D_1,\mathcal D_2)\right|_{{\rm max}\, \mathcal D_1  = {\rm max}\, \mathcal D_2}} {p(\mathcal D_1,\mathcal D_2)},
	\label{eq:tverde}
\end{equation}
where $\left.\bar p(\mathcal D_1,\mathcal D_2)\right|_{{\rm max}\, \mathcal D_1  = {\rm max}\, \mathcal D_2}$ is defined as the evidence when shifting $\Theta$ in likelihood $p(\mathcal D_2|\Theta)$ such that the maximum of $p(\Theta|\mathcal D_1)$ is at the same position in parameter space as the maximum of $p(\Theta|\mathcal D_2)$ (see section \ref{sub:measures_of_agreement} for more details). 

In the linear Gaussian model, shifting the parameter in the likelihood $p(\mathcal D_2|\Theta)$ is equivalent to changing the linear model for $\mathcal D_2$ to $F_2'(\Theta) = F_2^0 + M_2 (\Theta + \Theta')$. We have to choose $\Theta'$ such that the mean of the new posterior $\Theta_2'$ is equal to $\Theta_1$: 
\begin{equation}
	\Theta_2' = \Theta_1 + \Sigma_2 M_2^T C_2^{-1} (\mathcal D_2 - F_2'(\Theta_1)) \stackrel {!}{=} \Theta_1.
\end{equation}
Solving for $\Theta'$ and plugging it into \eqref{eq:tverde}, we find that $T$ evaluates to:
\begin{equation}
	\log T = \frac 1 2 (\mathcal D_2 - F_2(\Theta_1))^T K (\mathcal D_2 - F_2(\Theta_1)),
	\label{eq:logt}
\end{equation}
where
\begin{equation}
	K = C_2^{-1}M_2\Sigma_2(\Sigma_1 + \Sigma_2)^{-1}\Sigma_2M_2^TC_2^{-1}.
\end{equation}
The measure $T$ hence measures the difference between the data $\mathcal D_2$ and the prediction for the data from the posterior of $\mathcal D_1$. Averaging \eqref{eq:logt} over $p(\mathcal D_1, \mathcal D_2)$, we find that this measure is expected to vary around the following value for consistent datasets:
\begin{equation}
	\langle \log T \rangle = \frac d 2,
\end{equation}
where $d$ is the dimensionality of the parameter space $\Theta$.

Finally, we take a look at the measure by~\citet{MacCrann:2015ke} which is the p-value of the best-fit point of the joint analysis on the posteriors of the individual constraints. Without loss of generality, we consider the distribution of the best-fit point of the joint distribution as compared to the distribution of the posterior $p(\Theta|\mathcal D_1)$. In the linear Gaussian model the best fit point is simply given by the mean of $p_{\rm joint}$:
\begin{equation}
	\Theta_{\rm joint} = \Theta_1 + \Sigma_{\rm joint}M_2^T C_2^{-1}(\mathcal D_2 - F_2(\Theta_1))
\end{equation}
with $\Sigma_{\rm joint} = (\Sigma_1^{-1} + M_2^T C_2^{-1} M_2)^{-1}$ and $\Theta_1$ and $\Sigma_1$ being mean and covariance of $p(\Theta|\mathcal D_1)$. Under $p(\mathcal D_2|\mathcal D_1)$, $\Theta_{\rm joint}$ is expected to be normally distributed with mean $\Theta_1$ and covariance $\Sigma_1 - \Sigma_{\rm joint}$:
\begin{equation}
	\Theta_{\rm joint} \leftarrow \mathcal N(\Theta;\Theta_1,\Sigma_1 - \Sigma_{\rm joint}).
\end{equation}
Using the $p$-value of $\Theta_{\rm joint}$ under $p_1$ is hence a good approximation for the significance of the shift only if $\Sigma_{\rm joint} \ll \Sigma_1$, i.e. when the joint posterior is much tighter than the individual constraints, and is otherwise underestimating it.

\end{document}